\documentclass[aip,jap,preprint]{revtex4-1}

\usepackage[T1]{fontenc}        
\usepackage{geometry} 
\usepackage{graphicx}
\usepackage[english]{babel}   
\usepackage{amssymb}
\usepackage{epstopdf}
\usepackage{tikz}
\usepackage{pgfplots}
\usepackage{amsmath,amssymb,bm,times}
\usepackage{subfigure}

\begin{document}

\title{Elastic constants of polycrystalline steel evaluated with laser generated surface acoustic waves}

\author{D. Gasteau}
\email{damien.gasteau.etu@univ-lemans.fr}
\affiliation{LUNAM Universit\'es, CNRS, Universit\'e du Maine, LAUM UMR-CNRS 6613, Av. O. Messiaen, 72085 Le Mans, France.}
\affiliation{CEA Saclay DIGITEO Labs, F-91191 Gif-Sur-Yvette, France.}
\author{N. Chigarev}
\affiliation{LUNAM Universit\'es, CNRS, Universit\'e du Maine, LAUM UMR-CNRS 6613, Av. O. Messiaen, 72085 Le Mans, France.}
\author{L. Ducousso-Ganjehi}
\affiliation{CEA Saclay DIGITEO Labs, F-91191 Gif-Sur-Yvette, France.}
\author{V.~E. Gusev}
\affiliation{LUNAM Universit\'es, CNRS, Universit\'e du Maine, LAUM UMR-CNRS 6613, Av. O. Messiaen, 72085 Le Mans, France.}
\author{F. Jenson}
\affiliation{CEA Saclay DIGITEO Labs, F-91191 Gif-Sur-Yvette, France.}
\author{P. Calmon}
\affiliation{CEA Saclay DIGITEO Labs, F-91191 Gif-Sur-Yvette, France.}
\author{V. Tournat}
\email{vincent.tournat@univ-lemans.fr}
\affiliation{LUNAM Universit\'es, CNRS, Universit\'e du Maine, LAUM UMR-CNRS 6613, Av. O. Messiaen, 72085 Le Mans, France.}

\begin{abstract}
We report on a laser generated and detected surface acoustic wave method for evaluating the elastic constants of micro-crystals composing polycrystalline steel. The method is based on the measurement of surface wave velocities in many micro-crystals oriented randomly relative to both the wave propagation direction and the sample surface. The surface wave velocity distribution is obtained experimentally thanks to the scanning potentiality of the method and is then compared to the theoretical one. The inverse problem can then be solved, leading to the determination of three elastic constants of the cubic symmetry micro-crystals. Extensions of the method to the characterization of texture, preferential orientation of micro-crystals or welds could be foreseen.
\end{abstract}

\pacs{}
\keywords{Laser ultrasonics, NDE, Polycrystal}

\maketitle

\section{Introduction}

Most metals are, in the solid state, polycrystalline materials. They are composed of several crystallites of different sizes and orientations. Among those, austenitic steels are widely used in industry for their particular mechanical and thermal properties \cite{baddoo_stainless_2008,lo_recent_2009}. 

Considering the wide range of associated processes and final metallic products with various desired properties, the key steps towards fine control of high quality materials and processes could start with the identification of the micro-crystals orientations and geometries (for instance with micro-tomographic methods \cite{ludwig_three-dimensional_2009}), the evaluation of effective elastic properties by an average approach\cite{gnaupel-herold_calculation_1998,li_determination_1992}, the determination of the elastic constants of the constituents and the study of the grain boundary elastic properties (for instance with methods recently developed for surface breaking crack characterization \cite{ni_probing_2013,ni_crack_2012,mezil_all-optical_2011,dutton_near_2011,dixon_non-linear_2008}). 

Several techniques have been developed to characterize the microstructure and to assess surface material properties by using surface acoustic waves (SAW) or velocity as the contrast mechanism, as this varies with crystallographic orientation. Ultrasonic techniques that use surface acoustic waves (SAWs) for materials characterization include acoustic microscopy, conventional ultrasonics (contact or immersion), surface Brillouin scattering and laser ultrasonics. Among these techniques, contact techniques such as ultrasonic reflectivity and acoustic microscopy suffer from the couplant perturbation. Although, Brillouin scattering is noncontact, a high-quality sample surface is required and the signal to noise ratio tends to be rather poor due to the small proportion of Brillouin scattered photons resulting in extremely long measurement times. The laser generation and detection of ultrasound has been developed and proved to be a powerful tool to investigate SAWs \cite{li_determination_2012,zerr_elastic_2010,zerr_elastic_2012,hurley_time-resolved_2006,guilbaud_measurement_1999,edwards_enhancement_2011}.

Laser ultrasonics is an optical pump-probe technique allowing to analyze the propagation of bulk and surface waves without direct contact with the sample. For surface acoustic waves (SAW), the generated signal frequencies can be as high as several GHz, limited usually by the spatial structure of the laser spot focused on the surface (its diameter or spatial periodicity of fringes). Consequently, SAW of micro-scale wavelength can be generated and a typical propagation experiment can be carried out for source-detection distances of a few dozens of micrometers, sufficiently small to be achieved in a single micro-crystal. The elastic properties characterization of a micro-crystal is a problem with three unknowns which can be: the crystallographic orientation relative to the surface and the wave propagation direction, the micro-crystal elastic parameters, and the velocity of acoustic waves. For instance, measurement of slowness surface knowing the elastic parameters provides the orientation of the considered crystal \cite{li_determination_2012}, while an optimization on the measured phase velocities in an oriented sample can provide an estimation of elastic parameters \cite{reverdy_elastic_2001}. SAW waves in polycrystalline materials have also been used to monitor the propagation of waves through interfaces by time resolved measurements \cite{hurley_time-resolved_2006}.

We report in this article, a method to evaluate the elastic properties of the micro-crystals composing a polycrystalline sample. It is based on multiple measurements of the SAW velocities in randomly oriented micro-crystals, performed for the same direction propagation relative to the sample. As a result, different SAW velocities are extracted, providing a distribution of velocities, each having a probability of occurrence. The inversion of the problem consists of obtaining the three elastic constants of the cubic micro-crystals from the measured distribution of velocities. 

The paper is organized as follows. In Sec.\ref{METH}, the characteristics of the sample and of the experimental laser ultrasonics set-up are presented. In Sec.\ref{THEORY} the necessary theoretical elements relating the elastic constants of a cubic symmetry crystal arbitrarily oriented to the SAW velocities are recalled. In Sec.\ref{MONOC} typical experimental results are presented followed by a discussion about different causes of errors in Sec.\ref{PART_error}. Finally, the implemented method of inversion leading to the evaluation of the elastic parameters is detailed.

\section{Method and samples}
\label{METH}
	\subsection{Sample characteristics}
The studied sample is made of austenitic steel, with a cast composition $\mbox{Fe}_{68.34}\mbox{Cr}_{19.47}\mbox{Ni}_{9.64}$, where the indexes show the percentages in mass of the respective elements. From the complete mass percentage composition of the cast, the density of the sample is evaluated as $7628~kg/m^{3}$.  It is established that the grains present a cubic lattice symmetry and their typical size, following the NF A 04-102 norm, is comprised between 88 and 125 microns.

\begin{figure}[h!]
\begin{center}
\includegraphics[width=.7\columnwidth]{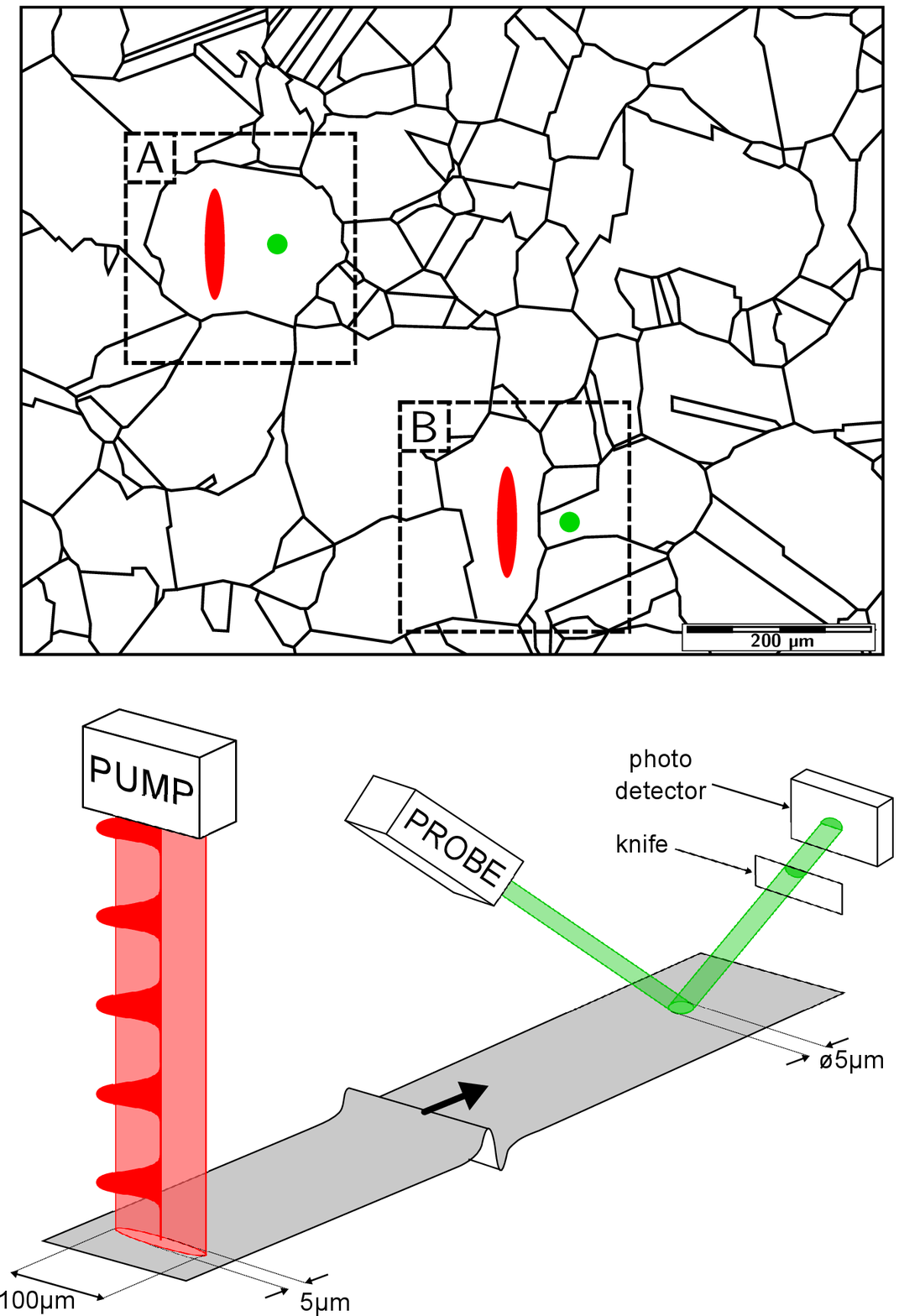}
\end{center}
\caption{(Color online) Representation of the sample surface (top) and schematics of the laser deflection method (bottom). The zones noted A and B are representative of the two possible dispositions of lasers spots compared to the crystallites boundaries.}
\label{FIG_surface_sample}
\end{figure}

Figure \ref{FIG_surface_sample} (top), obtained after identification of the boundaries from a picture of the surface taken by microscope, represents a part of the surface of the polycrystalline sample. The structural complexity of the material is visible and the locations designed by A and B are representative of the two categories of measurements done with the pump and probe laser spots. The pump is shown by the elongated spot while the probe is symbolized by the circular spot. Due to the heterogeneity in sizes and shapes of the micro-crystals, some of the measurements done while scanning the sample correspond to waves that propagates in a unique crystal (denoted by A in Fig.\ref{FIG_surface_sample}) but a non negligible part of the signals will correspond to the propagation of waves through several crystals interfaces as in the case B shown in Fig.\ref{FIG_surface_sample}. The influence of this kind of measurement is later detailed in Sec.\ref{PART_error}.

As presented in Sec.\ref{PART_setup}, the detection of the SAW pulse is based on the deflection of a laser beam reflected by the perturbed surface of the sample. To maximize this reflection and improve the quality of experimental signals, the surface of the studied sample is polished mechanically. As a result, the optical quality of the surface is ensured however the micro-crystalline structure is not visible anymore.

\subsection{Experimental setup}
	\label{PART_setup}
	
The generation of the SAW is performed via thermo-elastic effect induced by the absorption of a pump laser pulse represented by the elongated ellipse in Fig. \ref{FIG_surface_sample} (top). The pump laser has an optical wavelength of $1064~nm$ and the pulse duration is close to $0.75~ns$ with a repetition rate of $1~kHz$. The pump laser beam is focused on the sample surface into an streched ellipse (small diameter $\sim 5~\mu m$, big diameter $\sim 100~\mu m$) in order to generate plane surface waves. The second laser, used for probing the waves at the sample's surface, is shown by the circular spot in Fig.\ref{FIG_surface_sample}. This probe laser is continuous, at optical wavelength $532~nm$, and focused on a round spot of diameter $\sim 5~\mu m$ on the surface at a known distance from the pump laser spot. The reflected probe beam detects the local surface displacement via a beam deflection technique. From the pulses time of flight and the known distance between the pump and probe spots at the surface, the wave velocity can be deduced. Typical propagation distances range between 15 to 50$~\mu  m$, limited in the lower bound by the necessary distance of propagation to separate in time the different waves and in the higher bound by the SAW attenuation.

The proposed method of elastic properties determination is based on an analysis of the many local wave velocities that can be measured in such polycrystalline material. This ensemble of wave velocity values provides a wave velocity distribution which characteristics are sensitive to the elastic parameters of the material. The problem then lies in relating the wave velocity distribution to the elastic parameters of the material. 

\begin{figure}[h!]
\begin{center}
\includegraphics[width=.95\columnwidth]{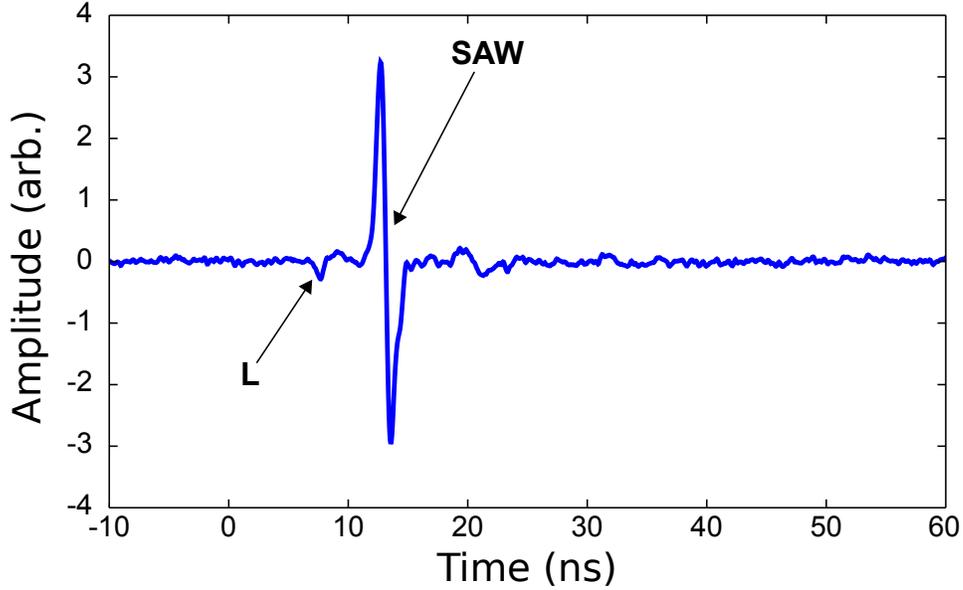}
\end{center}
\caption{Typical photoacoustic signal obtained by the pump-probe system.}
\label{FIG_time_signal}
\end{figure}

Figure \ref{FIG_time_signal} presents a typical temporal signal recorded by the pump-probe set-up. The SAW pulse is dominant and clearly identified. The longitudinal wave skimming along the surface (L) is also identified. As the detection procedure is mainly sensitive to the motions normal to the free surface, the Rayleigh wave, Generalized Rayleigh wave, the pseudo surface waves, and the skimming bulk waves \cite{favretto-cristini_elastic_2011} are mainly detected. In principle, there also exists two shear bulk waves, but their identification remains uncertain in the time signals because they do not present specific characteristics unlike the longitudinal wave, which is always the fastest, or the surface wave, which amplitude is an order of magnitude larger than the others.

\section{Surface waves in crystals}
	\label{THEORY}
	In this section, the main formulas leading to the eigenvalue problem to be solved to obtain the surface wave velocities for a semi infinite anisotropic crystal are recalled. The propagation direction of the surface wave is taken in the $x_1$ direction and the $x_2$ axis is oriented normally to the surface, pointing towards the solid bulk. The elastic properties of the material are described by the elastic stiffness tensor $c_{ijkl}$ ($i,j,k,l = 1,2,3$) which is usually reduced to a $6\times6$ matrix $c_{\alpha \beta}$ following the index reduction rules ($11\rightarrow1$, $22\rightarrow 2$, $33\rightarrow3$, $23$ or $32 \rightarrow4$, $13$ or $31\rightarrow5$, and $12$ or $21\rightarrow6$).

The problem is here expressed in terms of displacement and stress components as in Ref.\onlinecite{destrade_explicit_2001}, in order to obtain a matricial formulation of the problem similar to Refs. \onlinecite{fu_new_2002,mielke_uniqueness_2004}. The displacement components $u_n$ ($n=1,2,3$) along the direction $x_n$ can be written in a form of harmonic wave propagating along $x_1$ axis
\begin{equation}\label{eq_dep}
u_n(x_1,x_2)  = U_n (Kx_2) e^{\imath  (\omega t - K x_1)},
\end{equation}
where $U_n$ is a complex amplitude fonction, $K$ is the wavenumber, $\omega$ the pulsation, and $\imath = \sqrt{-1}$.

By the use of the stress-strain relationship,
\begin{equation}\label{eq_contdepray}
T_{ij} = c_{ijkl} \frac{\partial u_k}{\partial x_l},
\end{equation}
the motion equation,
\begin{equation}
\frac{\partial T_{ij}}{\partial x_j} = \rho \dfrac{\partial^2 u_i}{\partial t^2} ,
\end{equation}
and the stress expression $T_{ij} = K t_{ij}(Kx_2) e^{\imath(\omega t - K x_1)}$, 
the following differential matrix system is obtained,
\begin{equation}\label{eq_systglobal}
\begin{bmatrix}
A1 & Id \\
A3 - \rho v^2 Id & 0
\end{bmatrix}
\begin{Bmatrix}
U\\ t
\end{Bmatrix} = 
\begin{bmatrix}
B1 & 0 \\
B3 & Id
\end{bmatrix}
\begin{Bmatrix}
U' \\ t'
\end{Bmatrix},
\end{equation}
where $A1(i,j) = i c_{i21j}$, $A3(i,j) = c_{i11j}$, $B1(i,j) = i c_{i22j}$ and $B3(i,j) = -A1(j,i)$, $v=\omega/K$ is the SAW phase velocity, and the prime "$\ ' \ $" symbol denotes the partial derivative over $Kx_2$.

To determine the surface wave velocity, the condition of free stress at the surface is considered,
\begin{equation}
\begin{bmatrix}
A1 
\end{bmatrix}
\begin{Bmatrix}
U(0)
\end{Bmatrix} = 
\begin{bmatrix}
B1
\end{bmatrix}
\begin{Bmatrix}
U'(0)
\end{Bmatrix}.
\label{eq_CL}
\end{equation}

For given value of surface acoustic wave velocity $v$, the eigenproblem \eqref{eq_systglobal} is solved. The corresponding eigenvalues and eigenvectors are substituted in \eqref{eq_CL}. The value of $v$ for which \eqref{eq_CL} is fullfilled is the velocity of the surface waves. The solution for this particular problem has been demonstrated to be unique \cite{mielke_uniqueness_2004,fu_new_2002}. In some cases, this solution corresponds to the velocity of the Rayleigh wave \cite{lothe_existence_1976,barnett_consideration_1974} or to the velocity of the pseudo-surface wave \cite{farnell_properties_1970,lim_character_1969}.

The above theoretical procedure for SAW velocity calculation is then numerically implemented to obtain an histogram representation of the SAW velocities in random propagation directions and for randomly oriented crystal surfaces. In the following, a comparison is made between the numerically and experimentally obtained SAW velocity histograms. Then, an inversion procedure on the SAW velocity histograms is applied to estimate the elastic parameters of the tested material.

\section{SAW velocities for different crystal orientations}
\label{MONOC}

\begin{figure}[h!]
\begin{center}
\includegraphics[width=.95\columnwidth]{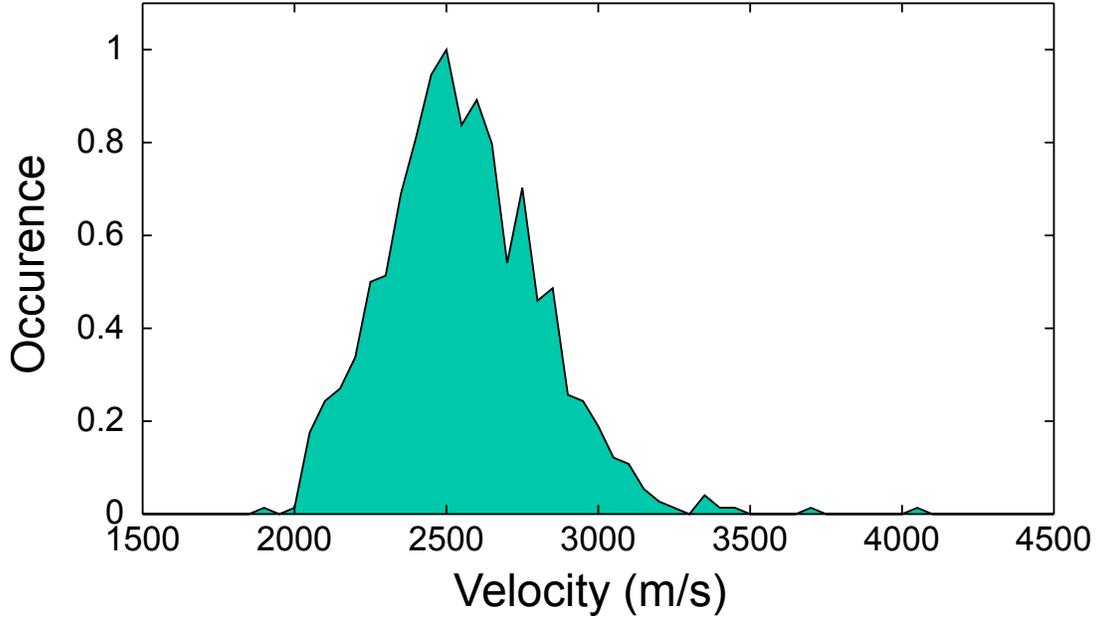}
\end{center}
\caption{(Color online) Normalized experimental histogram of the surface wave velocities in steel sample. Total of 850 measured velocities with bins of $50~m/s$ width.}
\label{FIG_hist_exp}
\end{figure}

Figure \ref{FIG_hist_exp} shows the wave velocity distribution (histogram) obtained experimentally after a scan of the sample surface. A thousand of wave velocity measurements have been performed by scanning the surface along 20 different lines of 50 positions each. All the positions of a line are spaced of 100 microns to insure that the spots are in different crystallites for each position. The propagation distance, between the pump and probe spots, is set to $42.8~\mu m$. This propagation distance is forced to be a multiple of $5.35~\mu m$ which is the experimental step of modulation of the distance between the two lasers spots. Approximately $15\%$ of the signals could not be exploited. Such insufficient signal quality could be explained by the poor local material surface optical quality, the signal-to-noise ratio, and difficulty in identifying wave pulses. The plot in Fig. \ref{FIG_hist_exp} is an histogram in which the velocity range is divided into bins of $50~m/s$ width.

Due to the number of velocities, this histogram is considered to be a good representation of all possible surface wave velocities that can be measured on the surface of the steel sample and their respective probability of occurrence. It is equivalent to a hypothetical measure of the surface wave velocities in different propagation directions and different cut planes of a single anisotropic crystal.

\begin{figure}[h!]
\begin{center}
\includegraphics[width=.95\columnwidth]{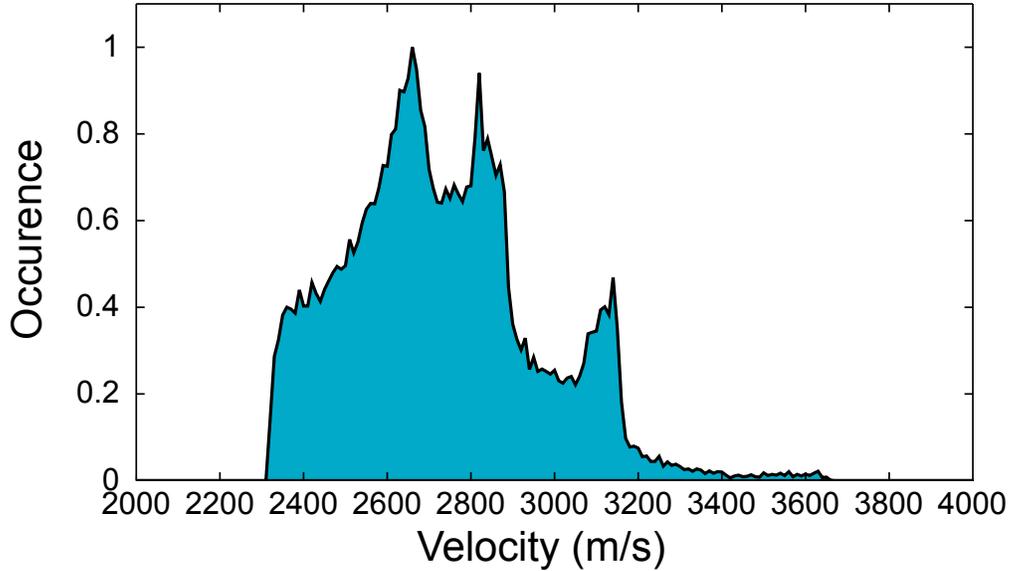}
\end{center}
\caption{(Color online) Numerically obtained histogram of surface wave velocities in iron ($c_{11} = 226~GPa$, $c_{12} = 140~GPa$, $c_{44} = 116~GPa$, $\rho = 7800~kg.m^{-3}$) for 50000 different orientations. The span of the velocities is here divided in bins with a width of $10~m/s$.}
\label{FIG_hist_th_Fe}
\end{figure}

Consequently, the theory presented in the Sec.\ref{THEORY} providing the surface wave velocity in one crystal depending of the surface orientation and propagation direction can be used to obtain a theoretical histogram. To do this, the cut plan is chosen randomly as well as the propagation direction and one velocity is theoretically obtained. This procedure is repeated many times with randomly chosen orientations to build the histogram. For instance, Fig.\ref{FIG_hist_th_Fe} shows the results obtained for iron ($c_{11} = 226~GPa$, $c_{12} = 140~GPa$, $c_{44} = 116~GPa$, $\rho = 7800~kg.m^{-3}$) for 50000 random orientations. With such a high number of calculated velocities, the space of orientation parameters is finely meshed. Each obtained velocity is calculated with a precision of $\pm 2.5~m/s$. This material is representative of the real sample under study in this article because the latter is made of steel, an iron alloy, and also shows a cubic lattice. 

Figure \ref{FIG_hist_th_Fe} clearly shows that the distribution of velocities is not constant nor gaussian. The same calculation has been made for nickel and copper, which are also cubic materials ; the results presented in the Appendix \ref{APP_other} show very similar qualitative features. A distribution with three peaks, a clear minimum and a tail at the largest velocities appear to be the characteristic features of cubic materials. Several of these features could be further used for solving more efficiently the inverse problem of elastic parameter evaluation from the velocity histogram, such as the value of the minimal velocity, as shown in Appendix \ref{APP_vmin}.

In the next section, limitations of the method are discussed. Then, the histogram representation is shown to be a specific signature of the elastic properties of a polycrystalline material, and an inversion procedure is applied to the velocity distributions, in order to evaluate the elastic constants of the corresponding material. 

\section{Limitations of the method}
\label{PART_error}
\subsection{Effect of Boundary crossing}
\label{PART_ERR}

While the automation of the set-up allows the acquisition of many signals, the optical quality of the surface is essential for the all-optical set-up to work in the best conditions and to loose the minimum of experimental data. This means that before the measurement, the surface of the sample must be polished and then the polycrystalline structure is not visible anymore. 

Consequently, while scanning the surface, the position of the spots relative to the crystal interfaces is unknown and the measurements can be expected to correspond to the two situations shown in Fig. \ref{FIG_surface_sample}. The case noted B shows the configuration of the laser spots where the detection spot is in another crystal than the generation spot. In this situation, the time of flight of the wave can be expected to correspond to a linear combination of the local velocities in the two crossed crystals. The probability to be in the situation B is directly linked to the ratio between the propagation distance and the size of the crystallites which can become a problem when crystal size is smaller and smaller.

On the other hand, with the actual set-up the propagation distance can not be reduced under around $15~\mu m$. At shorter propagation distance it is difficult to separate the different acoustic waves overlapping in time.

\begin{figure}[h!]
\begin{center}
\includegraphics[width=.95\columnwidth]{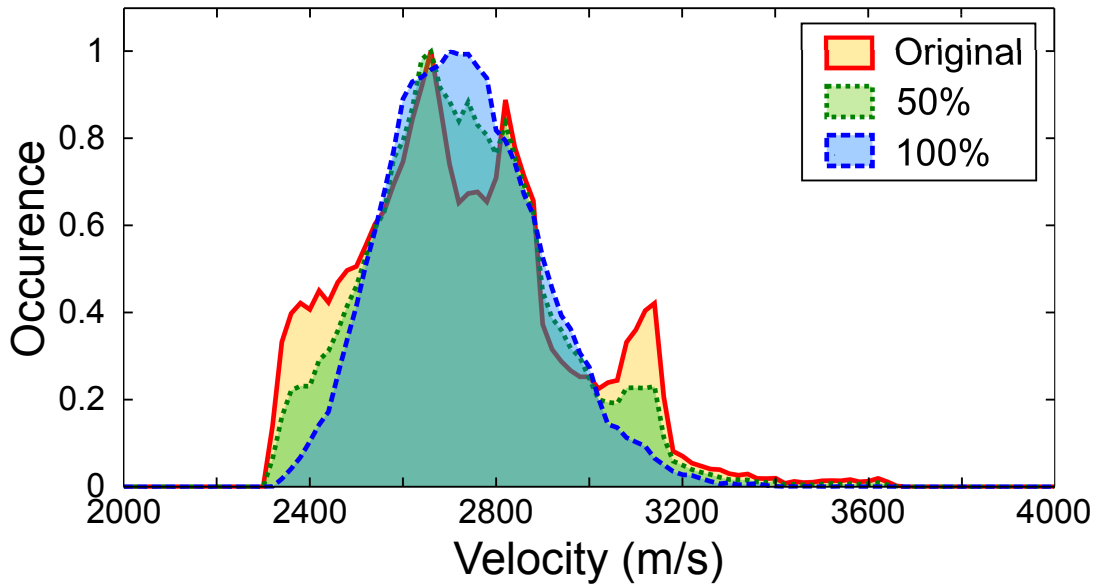}
\end{center}
\caption{(Color online) Effect of the error due to the boundary crossing on the reference histogram of iron. The legend indicates the percentage of boundary crossing estimated.}
\label{FIG_hist_err_bound}
\end{figure}

Figure \ref{FIG_hist_err_bound} shows the simulated histograms from the reference data on iron with several percentages of boundary crossing velocities (50 and 100). The histograms are created from the reference one in which a certain percentage of the velocities is replaced by a linear combination of two other velocities. It is visible that the more important is the boundary crossings, the less detailed is the histogram. But at a boundary crossing percentage of 50 percent, and consequently for smaller percentages, the main peak keeps its position which shows that the main features of the distribution are conserved even with an important proportion of boundary crossing. Even if the boundary crossing has a strong effect on the fast variations of the distribution, it does not modify the position of the peaks and the extremal values stay unchanged. 

Experimentally, it is not possible to find out wether a signal corresponds to the propagation of waves through an interface or not. Nevertheless, considering a propagation distance of $42.8~\mu m$ and the average size of the crystals of our sample between $88~\mu m$ and $125~\mu m$, the probability of boundary crossing should roughly not exceed $50\%$.

\subsection{Effect of time of flight estimation uncertainty}

\begin{figure}[h!]
\begin{center}
\includegraphics[width=.95\columnwidth]{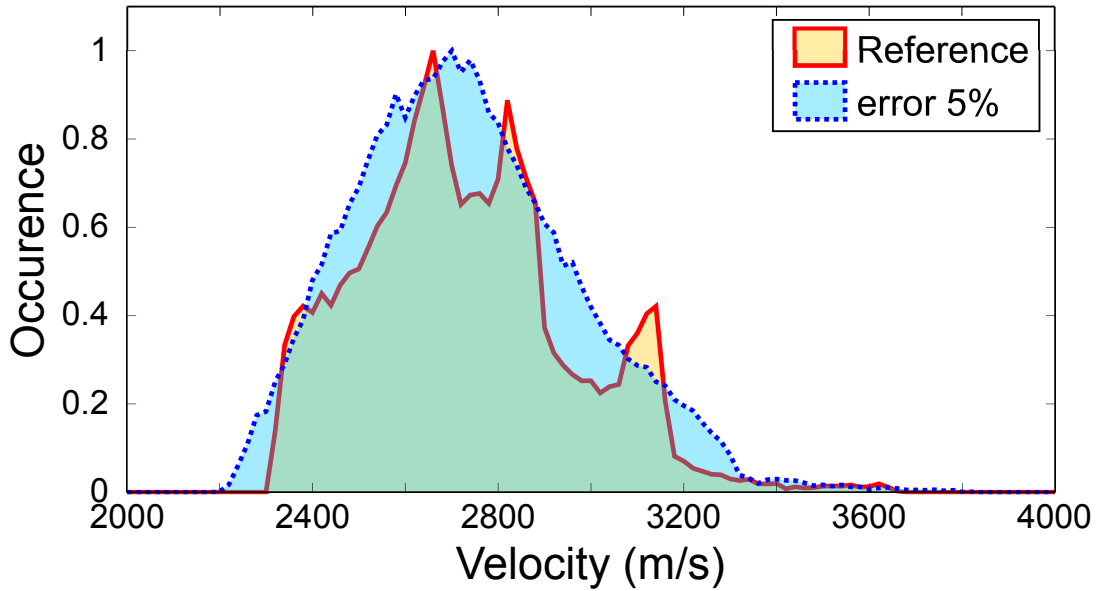}
\end{center}
\caption{(Color online) Effect of the experimental error on the velocities estimation (included in the interval $\pm 5\%$) on the histogram of iron.}
\label{FIG_hist_err_exp}
\end{figure}

Figure \ref{FIG_hist_err_exp} shows an estimation of the histogram of iron where the velocities values have been alterated to represent an experimental error spread over $\pm 5\%$ on the velocity estimations.

As the velocities are determined by measuring the time of flight and are inversely proportional to measured time delays, small errors on the data processing can bring an important deformation of the histogram as shown in Fig.\ref{FIG_hist_err_exp}. Then the impact of an error on the determination of the time of flight can have a larger influence than the error due to boundary crossing shown in Sec.\ref{PART_ERR}. Moreover, as the range of velocities is in the order of a few thousands of meters per seconds, a few percent of error implies $\sim100 m/s$ absolute error on the position and span of the histogram.

For this the pointing method of the arrival time of the different waves on the time signal is crucial. The experimental system is sensitive to the surface deformations, so the time signal actually shows the derivative of the surface displacement. To limit the ambiguity in time of flight determination, we choose to define the surface wave arrival as the time of zero crossing in the detected bipolar pulses. Note that at this time the displacement amplitude is maximum. 

\section{Inverse problem of elastic parameter evaluation}
\subsection{Inversion procedure}
	
As shown in Sec.\ref{THEORY}, from the elastic parameters of a material, the surface wave velocity in a chosen direction of propagation can be directly obtained. However, here the experimental situation is the opposite, the elastic parameters of the material are \textit{a priori} unknown, the orientations of crystals too and only the surface wave velocities can be assessed.

We start from the assumption that a given velocity distribution is characteristic of a set of elastic parameters, the inversion procedure is thus based on the minimization of a cost function between the histogram we have at our disposal and numerical histograms calculated for sets of known elastic parameters.

\begin{figure}[h!]
\centering
\includegraphics[width=.95\columnwidth]{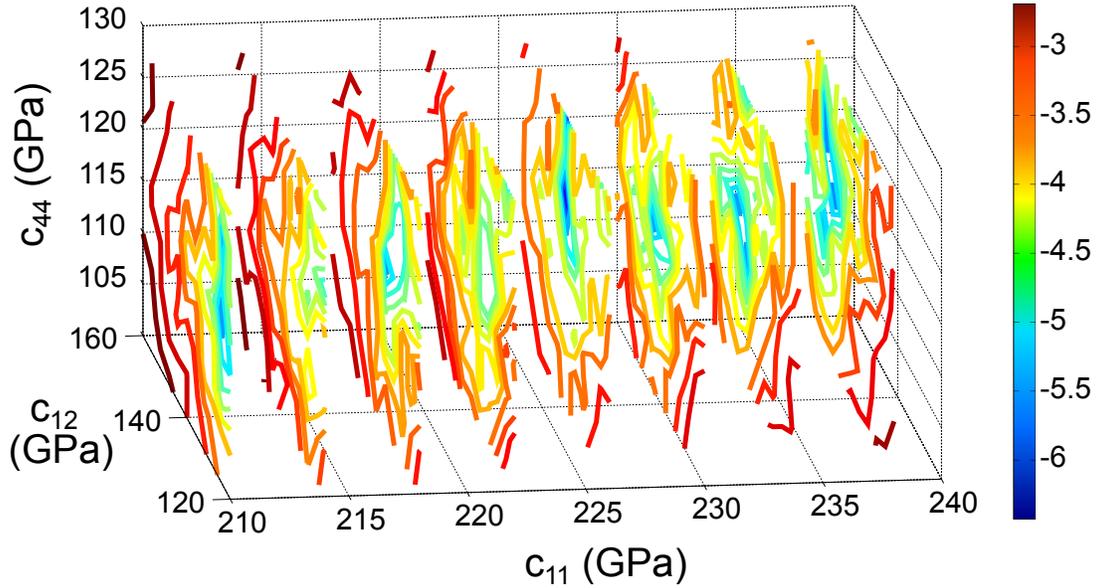}
\caption{Logarithm of the value of the cost function in the 3D elastic parameter space ($c_{11},c_{12},c_{44}$) for the histogram presented in Fig. \ref{FIG_hist_th_Fe}  with a space resolution of $4~GPa$. The visualization is based on a representation of isovalues of the cost function for each plan of $c_{11}$ constant.}
\label{FIG_err3D_th}
\end{figure}

To check the inversion procedure, the cost function value quantifying the error between the histogram presented in Fig.\ref{FIG_hist_th_Fe} and a test histogram obtained from three known values of ($c_{11}, c_{12}, c_{44}$) and for 1000 propagation directions is represented in the 3D space of parameter values ($c_{11}, c_{12}, c_{44}$) in Fig.\ref{FIG_err3D_th}.

The cost function in this inversion procedure has been chosen as follows. First, the same velocity bins are used for the two compared histograms. Second, for each velocity bin $i$ the occurence values are compared using the parameter $E_i$ defined as 
\begin{equation}
E_i = abs(1-\frac{A_i^{test}}{A_i^{ref}}),
\end{equation}
where $A_i^{test}$ and $A_i^{ref}$ are the value of the bin $i$ of respectively the test histogram and reference histogram.
Then, the cost function $CF$ can be expressed as 
\begin{equation}
CF = \overline{E} * \sigma (E),
\end{equation}
with $\overline{E}$ the mean value of $E$, $\sigma (E)$ the standard deviation of $E$ and $E = \left\lbrace E_1,E_2,...,E_N \right\rbrace$ the vector expressing all the errors $E_i$ on the different bins.

Figure \ref{FIG_err3D_th} is a representation of the cost function value between the reference histogram and the test histograms computed with triplets of elastic parameters ($c_{11}, c_{12}, c_{44}$) corresponding to their coordinates in space. A unique minimum of error is visible at the coordinates ($c_{11}=226~GPa, c_{12}=140~GPa, c_{44}=120~GPa$). The found minimum agrees well with the reference parameters ($c_{11}=226~GPa, c_{12}=140~GPa, c_{44}=116~GPa$). This is encouraging for the inversion procedure based on the comparison of histograms. The small difference on $c_{44}$ can be interpreted as the influence of the division of the velocity space in a finite number of bins. Also the fact that the comparison is done between a histogram based on 50000 values and histograms based on 1000 values could play a role on the precision of the results.

In the inversion procedure, the minimum of the cost function value is found through a minimization procedure starting from initial parameter values. The choice of initial parameters can have a strong influence on the computation time of the inversion procedure but we found no influence on the convergence of the solution. In Table \ref{TAB_inv_th} the results of six different runs of the minimization process (OPTIM) are shown.

\begin{table}[h!]
\centering
\begin{tabular}{|c|c|c|c|}
\hline 
 & $c_{11}~(GPa)$ & $c_{12}~(GPa)$ & $c_{44}~(GPa)$ \\ 
\hline 
Reference & 226 & 140 & 116 \\ 
\hline 
Starting param. & 200 & 150 & 100 \\ 
\hline 
OPTIM1 & 220.14 & 132.15 & 113.14 \\ 
\hline 
OPTIM2 & 224.45 & 135.77 & 113.49 \\ 
\hline 
OPTIM3 & 220.80 & 134.13 & 113.07 \\ 
\hline 
OPTIM4 & 222.87 & 133.77 & 111.94 \\ 
\hline 
OPTIM5 & 224.46 & 136.35 & 115.50 \\ 
\hline 
OPTIM6 & 216.85 & 129.39 & 115.51 \\ 
\hline 
Average & 221.6 & 133.6 & 113.8 \\
\hline
\end{tabular} 
\caption{Optimized elastic parameters for iron histogram.}
\label{TAB_inv_th}
\end{table}

The inversion procedure gives parameters close to the real ones (the error is in average less than 5\%) but slightly underestimated. Variations in the results arise from the limited number of random orientations used for the calculation of test histograms in each run.

\begin{figure}[h!]
\centering
\includegraphics[width=.95\columnwidth]{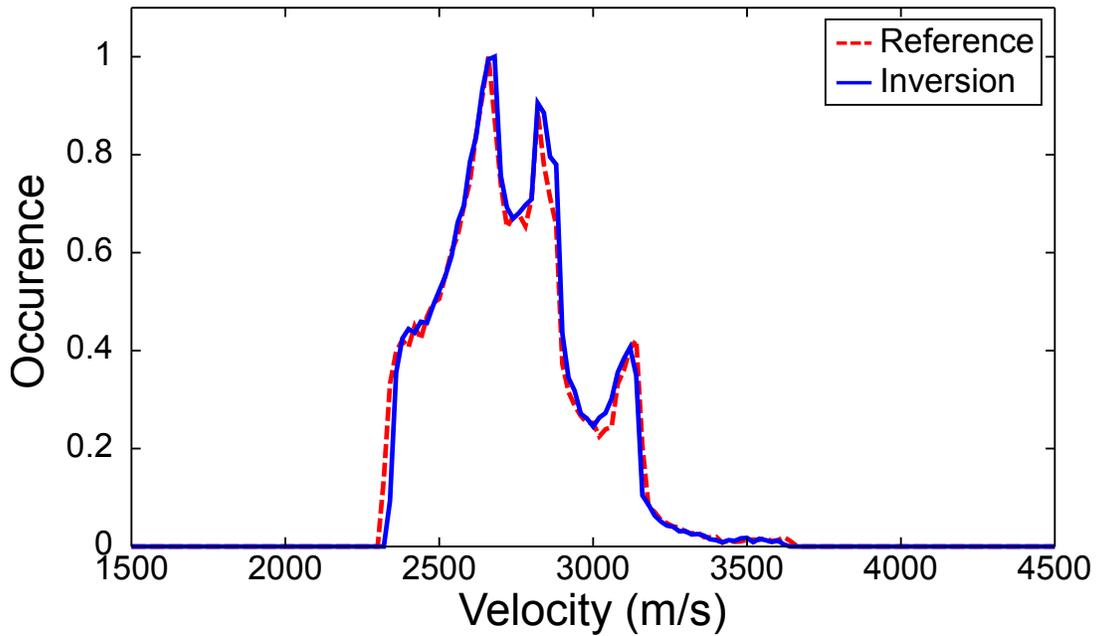}
\caption{(Color online) Histogram comparison between original  ($c_{11} = 226~GPa, c_{12} = 140~GPa, c_{44} = 116~GPa$) and reconstructed with the average of optimization results of Table \ref{TAB_inv_th} ($c_{11} = 221.6~GPa, c_{12} = 133.6~GPa, c_{44} = 113.8~GPa$).}
\label{FIG_comp_th}
\end{figure}

Taking the average values of the estimated elastic parameters presented in Table \ref{TAB_inv_th} a histogram with 50000 propagation directions can be calculated and compared to the original reference histogram. The result is shown in Fig.\ref{FIG_comp_th}. It can be seen that a few percent of error in the inversion procedure on the estimation of the elastic parameters does not correspond to an important change in the shape of the histogram.

\subsection{Inversion on experimental data}

\begin{figure}[h!]
\centering
\includegraphics[width=.95\columnwidth]{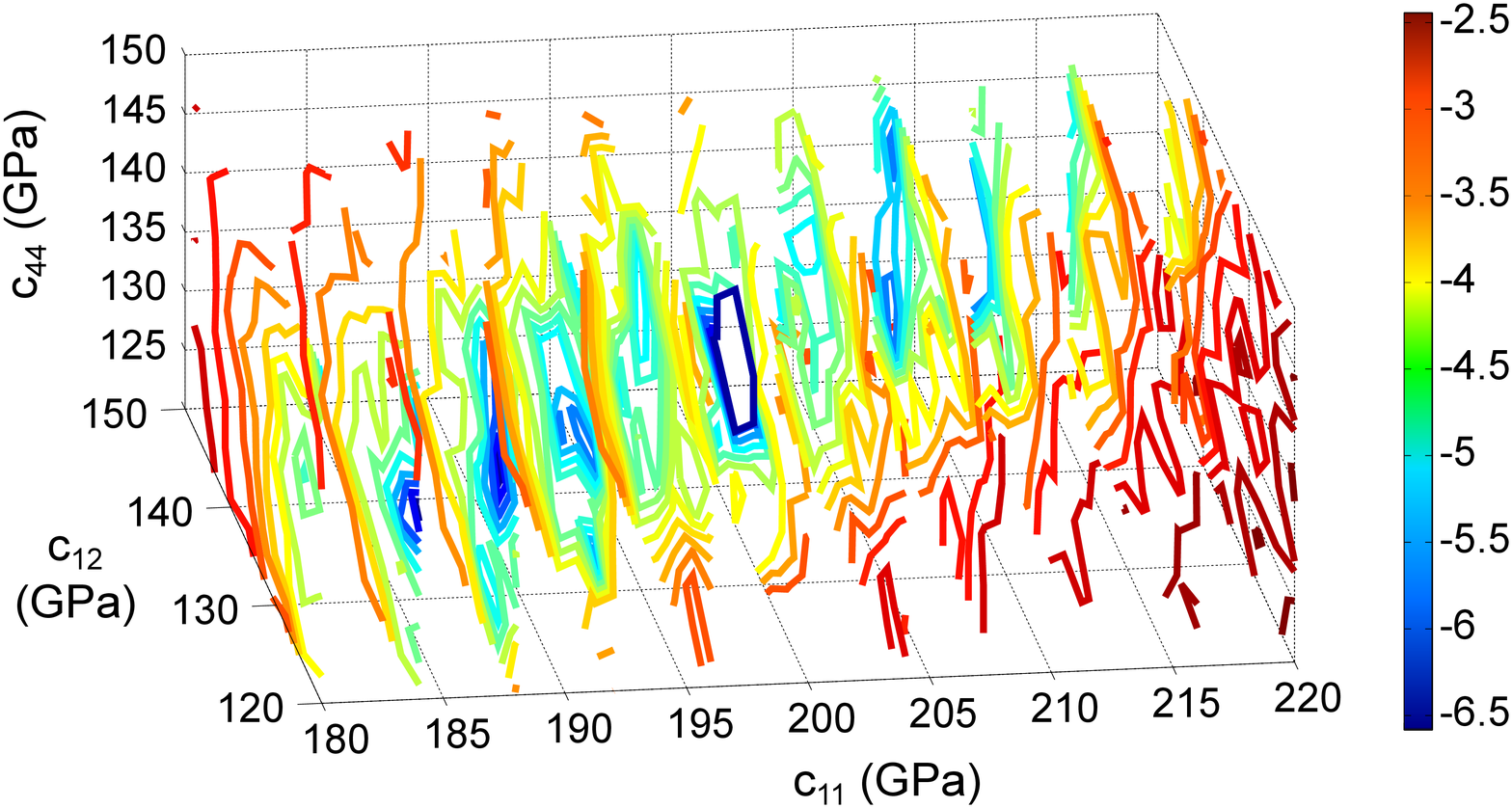}
\caption{(Color online) Logarithm of the value of the cost function in the 3D elastic parameter space ($c_{11},c_{12},c_{44}$) for the histogram presented in Fig. \ref{FIG_hist_exp}  with a space resolution of $4~GPa$. The visualization is based on a representation of isovalues of the cost function for each plan of $c_{11}$ constant.}
\label{FIG_err3D_exp}
\end{figure}

In order to check the existence of an unique set of elastic parameters as solution of a histogram representation of velocities, Fig.\ref{FIG_err3D_exp} shows the cost function value between a test histogram computed with a triplet of elastic parameters corresponding to its space coordinates and a reference histogram. Here the reference histogram is the one represented in Fig.\ref{FIG_hist_exp} which is deduced from a set of 850 velocities measured experimentally on the surface of the steel sample.

The zone of minimal error is more spread in the 3D space than for the theoretical case presented in Fig.\ref{FIG_err3D_th}. This can be explained by the error factors discussed in Sec.\ref{PART_error}. Nevertheless, the absolute minimum of error is identified, and corresponds to the set of parameters ($c_{11}=200~GPa, c_{12}=136~GPa, c_{44}=132~GPa$). This is close to the results of a recent study \cite{benyelloul_elastic_2013} where the composition of an austenitic steel sample is ($Fe_{62}Cr_{18.5}Ni_{18.5}$) and the parameters are estimated as ($c_{11}=200.4~GPa, c_{12}=129.3~GPa, c_{44}=125.8~GPa$). The set of parameters with the minimum error shows a good agreement with these values.

The inversion procedure is now modified to take in account a possible experimental error on the determination of the velocities. It is supposed that a realistic estimation of the experimental error belongs to the interval of $\pm 5\%$ on each velocity measured. The results of five runs of the inversion procedure are shown in Table \ref{TAB_inv_exp}.

\begin{table}[h!]
\centering
\begin{tabular}{|c|c|c|c|}
\hline
 & $c_{11}~(GPa)$ & $c_{12}~(GPa)$ & $c_{44}~(GPa)$ \\ 
\hline 
Starting param. & 200 & 120 & 100 \\ 
\hline 
OPTIM1 & 196.90 & 136.05 & 134.46 \\ 
\hline 
OPTIM2 & 198.31 & 137.24 & 140.60 \\ 
\hline 
OPTIM3 & 190.64 & 128.99 & 116.93 \\ 
\hline 
OPTIM4 & 189.51 & 126.30 & 121.77 \\ 
\hline 
OPTIM5 & 188.03 & 127.95 & 117.91 \\
\hline
Average & 192.7 & 131.3 & 126.3 \\
\hline
\end{tabular} 
\caption{Optimized elastic parameters for experimental histogram of austenitic steel.}
\label{TAB_inv_exp}
\end{table}

\begin{figure}[h!]
\centering
\includegraphics[width=1\columnwidth]{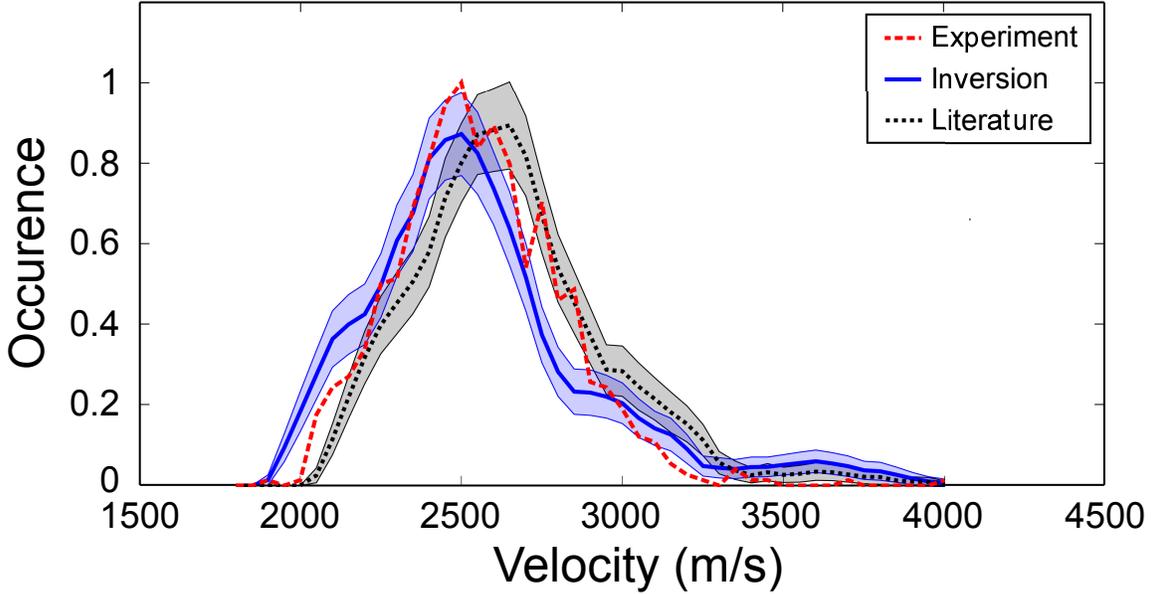}
\caption{(Color online) Histogram comparison between experimental (dashed red), inversion results, literature and their respective standard deviation translating the variations implied by the limited number of velocities considered.}
\label{FIG_comp_exp}
\end{figure}

Taking the average value of elastic parameters of the Table \ref{TAB_inv_exp} obtained by the inversion and the set of parameters from literature \cite{benyelloul_elastic_2013}, the histograms presented in Fig. \ref{FIG_comp_exp} are obtained. 

The velocities used to compute each numerical histogram in Fig.\ref{FIG_comp_exp} have been modified in order to simulate an experimental error of $\pm 5\%$. The lines named "Inversion" and "Literature" correspond to the average of 1000 histograms obtained from 850 velocity values with the parameters ($c_{11}=192.7~GPa, c_{12}=131.3~GPa, c_{44}=126.3~GPa$) and ($c_{11}=200.4~GPa, c_{12}=129.3~GPa, c_{44}=125.8~GPa$), respectively. The colored area around the average histograms corresponds to the standard deviation of the set 1000 histograms of 850 values considered. The Fig.\ref{FIG_comp_exp} shows that even with a relatively small number of velocity values, the numerical and experimental histograms can be matched with the inversion procedure. We also note that the inverted elastic parameters show good agreement with those of the literature.

\section{Conclusion}
A method to evaluate the elastic parameters of a polycrystalline material is reported, based on the analysis of multiple SAW velocities measured thanks to an all-optical pump-probe system. The surface of the studied steel sample is scanned and the measured local velocities of the SAW are used in the form of a distribution function. Considering a sample with random orientations of cubic crystallites of the same phase, the velocities that can be measured on its surface are analogous to the velocities assessed in different random orientations of a unique crystal. It is shown that this distribution is a signature of the anisotropic elastic properties of the crystallites without any prior information about the orientation of individual crystallites. The velocity distribution also shows a non trivial form, some peculiar velocities being measured more often than others. 

It is checked that a histogram representation is specific to a unique set of elastic parameters, and can be consequently used in an inversion procedure to estimate the corresponding elastic coefficients. The efficiency of the inversion procedure is shown to be sensitive to the quality of the recorded velocity histogram. Also the inherent uncertainties coming from the experiment are discussed as independent factors of error altering the efficiency and quality of the inversion procedure. The propagation of waves through an interface between micro-crystals and the error in the time of flight estimation are identified as the main factors of data deterioration. Also, to improve the convergence and precision of inversion results, an apriori information on the longitudinal skimming wave velocities could be used. For instance, the minimum value of longitudinal wave velocity is directly related to $c_{11}$ in a cubic crystal.

Further adaptations of this method could be directed towards the characterizations of preferential crystal orientations in a polycrystalline material such as in some casted steels or welds. The extension to the case of a non cubic symmetry is another possibility as well as acousto-elastic measurements based on this method for estimating the third order elastic constants of polycrystalline steels.

\appendix	
\section{Influence of the crystal material}
\label{APP_other}
\begin{figure}[h!]
\begin{center}
\includegraphics[width=\columnwidth]{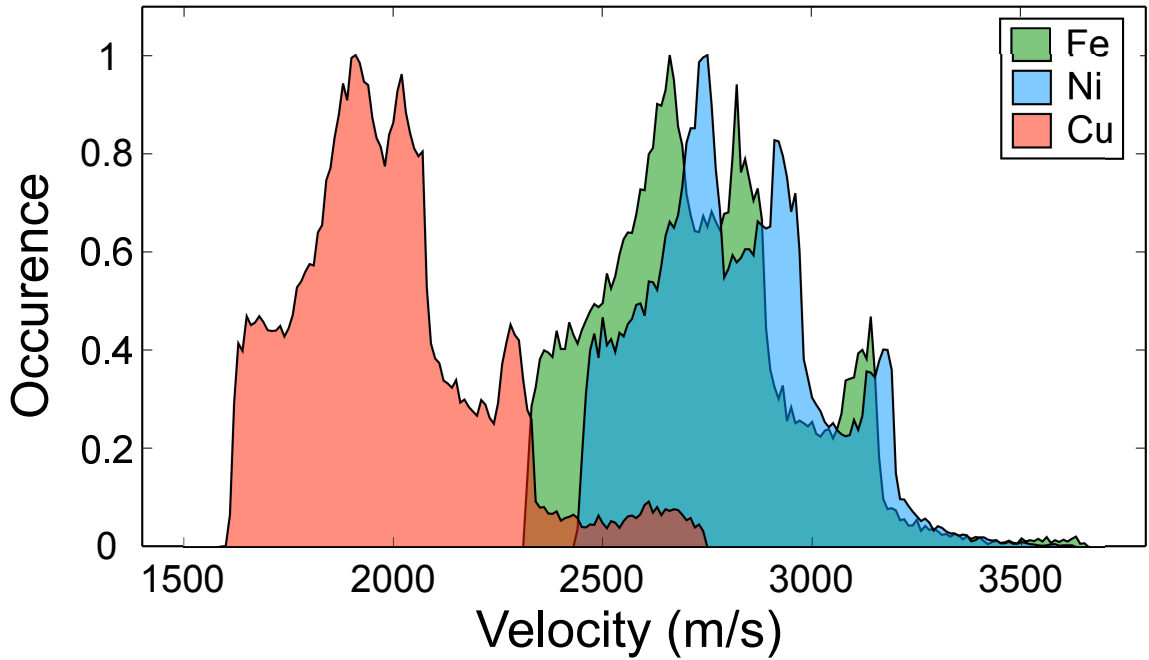}
\end{center}
\caption{(Color online) Theoretical histograms for three cubic materials. $c_{Fe} = (c_{11} = 226~GPa, c_{12} = 140~GPa, c_{44} = 116~GPa, \rho = 7800~kg/m^3)$, $c_{Cu} = (c_{11} = 168.4~GPa, c_{12} = 121.4~GPa, c_{44} = 75.2~GPa, \rho = 8960~kg/m^3)$, $c_{Ni} = (c_{11} = 261~GPa, c_{12} = 151~GPa, c_{44} = 130.9~GPa, \rho = 8902~kg/m^3)$}
\label{FIG_3hist}
\end{figure}

Figure \ref{FIG_3hist} shows examples of numerical histograms for the SAW velocities from 50000 random propagation directions for different materials, iron, nickel, and copper. All three materials have a cubic crystalline structure but different elastic parameters. While the global shapes of the histograms are similar, their positions on the velocity axis are clearly different.

\section{Evaluation of the minimum SAW velocity of an histogram}
\label{APP_vmin}
The minimum value of surface wave velocity has been identified in the plane $(110)$ and the propagation direction $(-110)$. In this orientation the velocity is a solution of the third order polynomial $aX^3+bX^2+cX+d = 0$, where $X = \rho v^2$ and
\begin{equation}
\begin{array}{l}
a = \frac{1}{C_{66}}-\frac{1}{C_{11}},\\
b = \frac{2}{C_{66}} \left(  \frac{C_{12}^2}{C_{11}} - C_{11} \right), \\
c =  \frac{1}{C_{66}} \left(  \frac{C_{12}^2}{C_{11}} - C_{11} \right)^2 - 2 \left(  \frac{C_{12}^2}{C_{11}} - C_{11} \right), \\
d = -\left(  \frac{C_{12}^2}{C_{11}} - C_{11} \right).
\end{array}
\end{equation}

with
\begin{equation}
\begin{array}{l}
C_{11} = \frac{c_{11}+c_{12}+2c_{44}}{2},\\
C_{12} = \frac{c_{11}+c_{12}-2c_{44}}{2},\\
C_{66} = \frac{c_{11}-c_{12}}{2}.
\end{array}
\end{equation}

The minimum velocity can be directly calculated knowing that the roots of a third order polynome can be expressed
\begin{equation}
\label{eq_solpoly}
x_k = \frac{-1}{3a} \left( b + u_k C + \frac{\Delta_0}{u_k C} \right)
\end{equation}
with
\begin{equation}
\begin{array}{l}
u_1 = 1,~~ u_2 = \frac{-1 + i \sqrt{3} }{2}, ~~u_3 = \frac{-1 - i \sqrt{3} }{2} \\
C = \sqrt[3]{\frac{\Delta_1 + \sqrt{\Delta_1^2 - 4 \Delta_0^3}}{2}} \\
\Delta_0 = b^2-3ac,\\
\Delta_1 = 2b^3 - 9abc + 27 a^2d, ~~\Delta_1^2 - 4 \Delta_0^3 = -27a^2 \Delta, \\
\Delta = 18 abcd - 4 b^3d + b^2c^2 - 4 ac^3 - 27a^2d^2.
\end{array}
\end{equation}

Considering the real solution of the polynomial expression, the following table shows the results of \eqref{eq_solpoly} where $v_{min}^{hist}$ is the minimum value of velocity found by calculating an histogram of 900 different orientations and $v_{min}^{th}$ the result given by the expression of the minimal velocity.
\begin{table}[h!]
\centering
\begin{tabular}{|c|c|c|c|c|c|c|c|c|}
\hline 
Material 							    & Fe 	    & Ni 			& Au 		& Cu 		    \\
\hline
$c_{11}~(GPa)$ 					& 226 		& 261		& 202		& 168.4		\\
$c_{12}~(GPa)$						& 140 		& 151		& 169.7	& 121.4		\\
$c_{44}~(GPa)$						& 116		& 130.9	& 46.02	& 75.2			\\
$\rho~(kg.m^{-3})$ 			& 7800		& 8902		& 19300	& 8960			\\
\hline
$v_{min}^{hist}~(m/s)$ 		& 2320 	& 2445 	& 915 		& 1615 		\\
$v_{min}^{th}~(m/s)$			& 2317.5	& 2443.8	& 908.5	& 1606.8		\\
error (\%) 								& 0.1 		& 0.04 		& 0.7 		& 0.5\\
\hline
\end{tabular} 
\caption{Table comparing the minimum values of surface acoustic wave velocities for different cubic materials}
\label{table_vmin}
\end{table}

\end{document}